# Factors Influencing the Adoption of Cloud Incident Handling Strategy: A Preliminary Study in Malaysia

*Full paper*

**Nurul Hidayah Ab Rahman**
Information Assurance Research Lab,
University of South Australia, GPO Box
2471, Adelaide, SA 5001, Australia
abyny002@mymail.unisa.edu.au

**Kim-Kwang Raymond Choo**
Information Assurance Research Lab,
University of South Australia, GPO Box
2471, Adelaide, SA 5001, Australia
raymond.choo@unisa.edu.au

## Abstract

This study seeks to understand the factors influencing the adoption of an incident handling strategy by organisational cloud service users. We propose a conceptual model that draws upon the Situation Awareness (SA) model and Protection Motivation Theory (PMT) to guide this research. 40 organisational cloud service users in Malaysia were surveyed. We also conduct face-to-face interviews with participants from four of the organisations. Findings from the study indicate that four PMT factors (Perceived Vulnerability, Self-Efficacy, Response Efficacy, and Perceived Severity) have a significantly influence on the adoption of cloud incident handling strategy within the organisations. We, therefore, suggest a successful adoption cloud incident handling strategy by organisational cloud service users involves the nexus between these four PMT factors. We also outline future research required to validate the model.

## Introduction

Cloud computing has been highlighted as a key initiative in Malaysia's Information and Communication Technology (ICT) strategic plan with a number of projects funded to advance the development and deployment of cloud services among government agencies. In 2012, for example, the Multimedia Development Corporation (MDeC) of Malaysia launched a "Small Medium Enterprise (SME) Cloud Computing Adoption Programme", and provided financial incentives to increase cloud adoption within the country (Multimedia Development Corporation 2012). In a more recent study by Asia Cloud Computing Association (ACCA), Malaysia ranks eighth in the Asia Pacific Cloud Readiness Index which measures 13 other countries cloud readiness (ACCA 2014).

Migrating to cloud is not, however, without challenges, particularly for organisations and government agencies dealing with sensitive information such as those of their users and citizens respectively. Cloud and related security threats are real, as explained by scholars (Choo 2010; Quick et al. 2014) and highlighted in recent high profile incidents (European Parliament 2014; Federal Bureau of Investigation 2014). Consequences range from national security breaches to loss of business (e.g. due to civil litigation and loss of competitive advantage) to embarrassment, and can result in substantial losses, both direct (e.g. financial) and indirect (e.g. reputational), to an organisation. A sound, proactive, and preferably evidence-based, strategy to handle and respond to a security incident must be taken into account when migrating to the cloud environment. This was also highlighted by the Cloud Security Alliance (2011). An incident handling strategy generally includes pre-incident, incident detection, incident analysis, containment, eradication and recovery (i.e. incident response) (Cichonski and Scarfone 2012; Killcrece 2003).

There are, however, challenges in adapting a "traditional" incident handling strategy for the cloud due to the nature of the cloud infrastructure as explicated in a recent survey (Ab Rahman and Choo 2015a). For example, Cloud Service Users (CSUs), including organisational CSUs, may not have easy access to information pertaining to a security breach or are unaware of a security breach involving their data, may have limited insight and knowledge about the underlying cloud architecture, and the uncertainty in roles and responsibilities for incident handler (Grobauer and Schreck 2010; Hooper et al. 2013; Lenkala et al. 2013; Loske et al. 2014). The challenges are compounded by the (significant) variation between organisational CSUs and CSPs, as well as the underlying cloud architecture and deployment model (Ab Rahman and Choo 2015b). Despite the importance of incident handling, a survey of 23 European countries by the European Union Agency for Network and Information Security (2013) suggested that incident handling remains a concern for the majority of the surveyed countries. It is, therefore, necessary to understand what factors influence the adoption of cloud incident handling strategy by organisational CSUs – the objective of this study.

To address the research objective, we propose a conceptual model which draws upon the Situational Awareness (Taylor 1990) Model and the Protection Motivation Theory (Maddux & Rogers 1983). This model is used as the underlying theoretical lens, and is described in the next section. We then describe the instrument design and data collection strategy, followed by the findings and avenues for future research.

## Conceptual Model and Hypotheses

Situational Awareness (SA), the first underlying concept in this study, was first used in the aviation industry to collect different environmental data to facilitate dynamic decision-making (Taylor 1990). The analysis of a situation encompasses three increasing key levels introduced in a latter seminal study by Endsley (1995), namely: *perception*, *comprehension*, and *projection*. As discussed by Endsley (1995), perception forms a basic set of knowledge of significant elements in the environment by obtaining information such as status, attributes, and dynamics of certain elements. Comprehension is regarded as people developing an understanding or awareness of the environment's elements by combining, interpreting, storing, and retaining information. This will help one to subsequently perceive a situation and assess the level of knowledge, which leads to the Projection level.

The second underlying concept of this study is the Protection Motivation Theory (PMT), an extension of the Health Belief Model (Rogers 1975). PMT sets out to understand and clarify *fear appeals* in order to change *attitudes* and *behaviours* and overcome fear (Maddux & Rogers 1983). There are two key components – threat appraisal and coping appraisal – in PMT. Threat appraisal refers to the assessment of the risk level posed by a threatening event, and consists of the following factors:

i. Perceived vulnerability — individuals were believed to vary widely in their acceptance of personal susceptibility to a condition; and
ii. Perceived severity — fears that a person has regarding the significance of a threat.

The second component is coping appraisal, which is regarded as the assessment of one's own ability to handle and prevent loss occurring from the threat (Ifinedo 2012; Maddux and Rogers 1983). The coping appraisal comprises three factors:

i. Self-efficacy — the degree of confidence in an individual's ability to execute an action successfully (Bandura 1977; Maddux and Rogers 1983);
ii. Response efficacy — beliefs regarding the effectiveness of the various actions available in reducing the threat. It is expected that an individual would not be expected to accept the recommended action unless it was perceived as feasible and effective (Maddux & Rogers 1983; Rogers 1975); and
iii. Response cost — the perceived opportunity costs in terms of money, time and effort expended when implementing the recommended behaviour (Ifinedo 2012). Some potential negative aspects (e.g. expensive, time-consuming, and dangerous) may act as barriers to a recommended action.

The concept of SA has gained traction in the information security discipline from both technological and management perspectives. Motivated to provide effective Honeynet data analysis, for example, Yegneswaran et al. (2005) and Bing et al. (2012) demonstrated that SA is useful to facilitate one in classifying and summarising a dataset in a network security context. Zeng et al. (2014) and Webb et al. (2014) demonstrated the potential for using cyber security SA as a basis for analysing the correlation state machine data to obtain up-to-date cyber security information and knowledge, and the basis of an information security risk management cycle, respectively. In the context of incident handling, Skopik et al. (2012) and Murray and Ruefle (2014), applied SA in the incident response cycle of an Cyber Attack Information System, and as an organisational requirement in decision-making to prevent, detect and respond to information security threats and risks, respectively.

PMT has also been studied in the information security (IS) discipline. Examples of study that seek to understand whether PMT factors can influence users' security behaviour include those that examined IS behaviour (Claar and Johnson 2012; Johnston and Warkentin 2010; Ng et al. 2009), information seeking behaviour (Wang et al. 2012), and IS policy compliance (Herath and Rao 2009; Ifinedo 2012; Siponen et al. 2014; Vance et al. 2012). In this study, we adapted the SA model and PMT factors in our conceptual model in order to measure the level of incident handling strategy adoption by organisational CSUs. In our model, the five PMT factors are grouped under the three levels of SA (Perception, Comprehension, Projection; see Table 1). The model also consists of five independent variables, namely: Perceived Vulnerability, Response Efficacy, Response Cost, Self-Efficacy, and Perceived Severity; and one dependent variable – i.e. User's incident handling strategy adoption (see Figure 1).

| SA Level | PMT Factors |
|---|---|
| Perception | Perceived vulnerability (PVUL) |
| Comprehension | Self-efficacy (SEF); Response efficacy (REF); Response cost (RCO) |
| Projection | Perceived severity (PSEV) |

**Table 1: SA level and the relevance PMT factors**

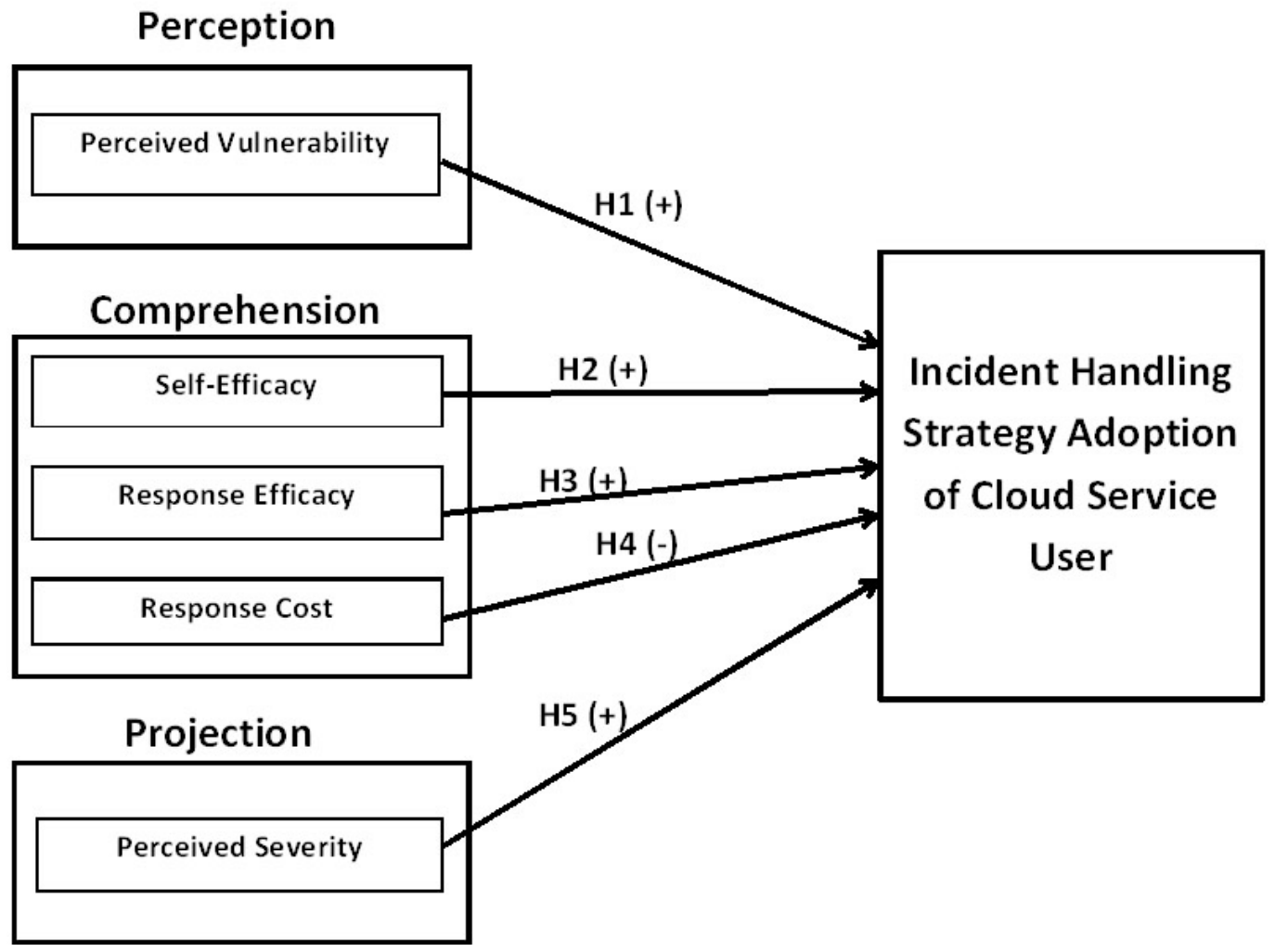

**Figure 1: Conceptual Model (integrating the Situation Awareness Model and Protection Motivation Theory)**

Perception involves the gathering of information, which creates knowledge of perceived cloud computing security and privacy threats such as potential threat actors, attack vector and target. It is accepted that cloud computing can introduce new security, privacy and trust risks and challenges, which reinforce the need for secure management of data (particularly in SaaS), privacy rules enforcement, and trust in CSP to do the right thing (Choo 2011, 2014). At this level, Perceived Vulnerability (PVUL) is the relevant factor and is defined as the CSU's perception of a security threat to the cloud's deployment. This will then lead to the adoption of an incident handling strategy.

**Hypothesis 1 (H1)**: Perceived vulnerability to incident occurrence is positively associated with the adoption of an incident handling strategy.

Comprehension at the second level enables CSUs to understand and assess the current security and privacy risks, as well as the benefits and potential challenges of adopting the incident handling strategy. Self-Efficacy (SEF), Response Efficacy (REF), and Response Cost (RCO) are, therefore, the relevant factors that need to be implemented at the Comprehension level. SEF refers to the level of CSUs' confidence to implement an incident handling strategy. REF is concerned with the perceived benefits of adopting incident handling strategy, and RCO refers to the inconvenience and other barriers in the adoption and implementation of an incident handling strategy. Examples of barriers including lack of support from senior management (Mitropoulos et al. 2006; Shedden et al. 2010; West-brown et al. 2003), and overly complicated procedures (Humaidi and Balakrishnan 2013). The hypotheses developed for these three factors are as follows:

**Hypothesis 2 (H2):** Self-efficacy is positively related to the CSP's incident handling strategy.
**Hypothesis 3 (H3)**: Response efficacy of incident handling in an organisation is positively related to the adoption of an incident handling strategy.
**Hypothesis 4 (H4)**: Perceived response costs of incident handling strategy are negatively related to its adoption.

At the Projection level, assessments from the earlier levels inform CSU's future incident handling strategy implementation. For instance, CSU would learn from previous implementation of an inefficient incident handling system (Bojanc 2013). Past studies have also indicated that perceived severity has a significant influence on user security behaviour (Ifinedo 2012; Johnston and Warkentin 2010; Liang and Xue 2010; Ng et al. 2009). Therefore, we regard perceived severity (PSEV) as a relevant influencing factor for a CSU to adopt an incident handling strategy.

**Hypothesis 5 (H5)**: Perceived severity positively affects CSUs' adoption of an incident handling strategy.

## Research Methodology

### *Instrument Design*

A questionnaire was designed to collect data for this study (see Table 2), which was divided into four sections. The first part seeks general information on the adoption of cloud services in the participating organisation. The second part seeks information pertaining to the participant's perceptions of cloud security and privacy threats. The third part is concerned with the participant's perceptions of information security incident handling policy in their organisation, and the fourth part solicits responses about participant's perceptions of their CSP information security incident handling strategies.

| SA Level | Factor | Item text | Measure | Sources |
|---|---|---|---|---|
| Perception | PVUL | What is the likelihood of the following cloud specific threats facing your organisation?<br>PVUL1 – Data breaches by insider<br>PVUL2 – Data breaches by outsider<br>PVUL3 – Data loss by insider<br>PVUL4 – Data loss by outsider<br>PVUL5 – Account or service traffic hijacking by insider<br>PVUL6 – by outsider | Most likely (1) to Not at all (5) | (Cloud Security Alliance 2013); and Self-developed |
| | | PVUL7 – Insecure software interfaces<br>PVUL8 – Denial of service<br>PVUL9 – Cloud service abuse by insider<br>PVUL10 – Cloud service abuse by outsider<br>PVUL11 – Inadequate due diligence<br>PVUL12 – Shared technology vulnerabilities | | |
| | | What is the likelihood of the following threat vectors being exploited to target your organisation's cloud services?<br>PVUL13 – People<br>PVUL14 – Process<br>PVUL15 – Technologies | Very high (1) to Very low (5) | Self-developed |
| | CUE | CUE1 – The content is updated regularly.<br>CUE2 – My organisation organises regular workshop or seminar on information security and/or incident handling. | Strongly agree (1) to Strongly disagree (5) | Self-developed |
| | GEN | GEN1 – My organisation's information security incident handling policy is compliant with existing security management and incidents handling standard (e.g. ISO/IEC 27001:2013 and ISO/IEC 27035:2011).<br>GEN2 - The workshop or seminar increases my awareness on information security and/or incident handling. | Strongly agree (1) to Strongly disagree (5)<br>Least aware (1) to Mostly aware (5) | Self-developed |
| Comprehension | SEF | SEF1 — My organisation provides in-house technical support to handle information security incident.<br>SEF2 — My organisation provides adequate in-house staff to handle information security incident.<br>SEF3 — My organisation provides adequate budget to handle information security incident. | Strongly agree (1) to Strongly disagree (5) | Self-developed |
| | REF | REF1 — The policy is adequate to handle non-cloud related information security incident(s) effectively.<br>REF2 — The policy is adequate to handle cloud-related information security incident(s) effectively.<br>REF3 — Based on my organisation's experiences with the Cloud Service Provider, my organisation is confident with the Cloud Service Provider's capability to handle an information security incident. | Strongly agree (1) to Strongly disagree (5) | Self-developed |
| | RCO | What are the barriers to an effective information security incident handling policy | Strongly agree (1) to Strongly | (Killcrece et al. 2003); and |
| | | implementation at your organisation?<br>RCO1 - Incident handling procedures are too complex to be executed<br>RCO2 - Lack of support from top management<br>RCO3 - Lack of budget and other resources | disagree (5) | Self-developed |
| Projection | PSEV | What are the consequences of an ineffective information security incident handling policy to an organisation (i.e. cloud service user)?<br>PSEV1 - Financial loss<br>PSEV2 - Reputation damage<br>PSEV3 - Legal implications | Strongly agree (1) to Strongly disagree (5) | (Killcrece et al. 2003); and Self-developed |

**Table 2: Measurement Items**

### *Data Collection*

A self-administered online survey was distributed to 100 organisational CSUs in Malaysia, which were identified in collaboration with MDeC. We had 40 responses, one from a different organisation. We also conducted face-to-face interviews with four participants who had indicated on the questionnaire survey that they would like to participate in the follow-up interview. The interviews were conducted in the Malay language, and to ensure anonymity, the interviewees are referred as I1, I2, I3, and I4 in this paper (see Table 3).

| Industry Type | Job Role | |
|---|---|---|
| | IT Security Expert | Cloud Project Manager |
| Financial institution | I1 | |
| Education | | I2 |
| ICT Services | I3 | |
| Audit, Tax, and Advisory Services | I4 | |

**Table 3: Interview Participants' Information**

## Data Analysis and Result

### *Demographic Analysis*

As can be seen in Figure 2, the three main participating industry types were government agencies (30%), educational institutions such as universities (25%) and information and culture organisations such as IT service providers (22%).

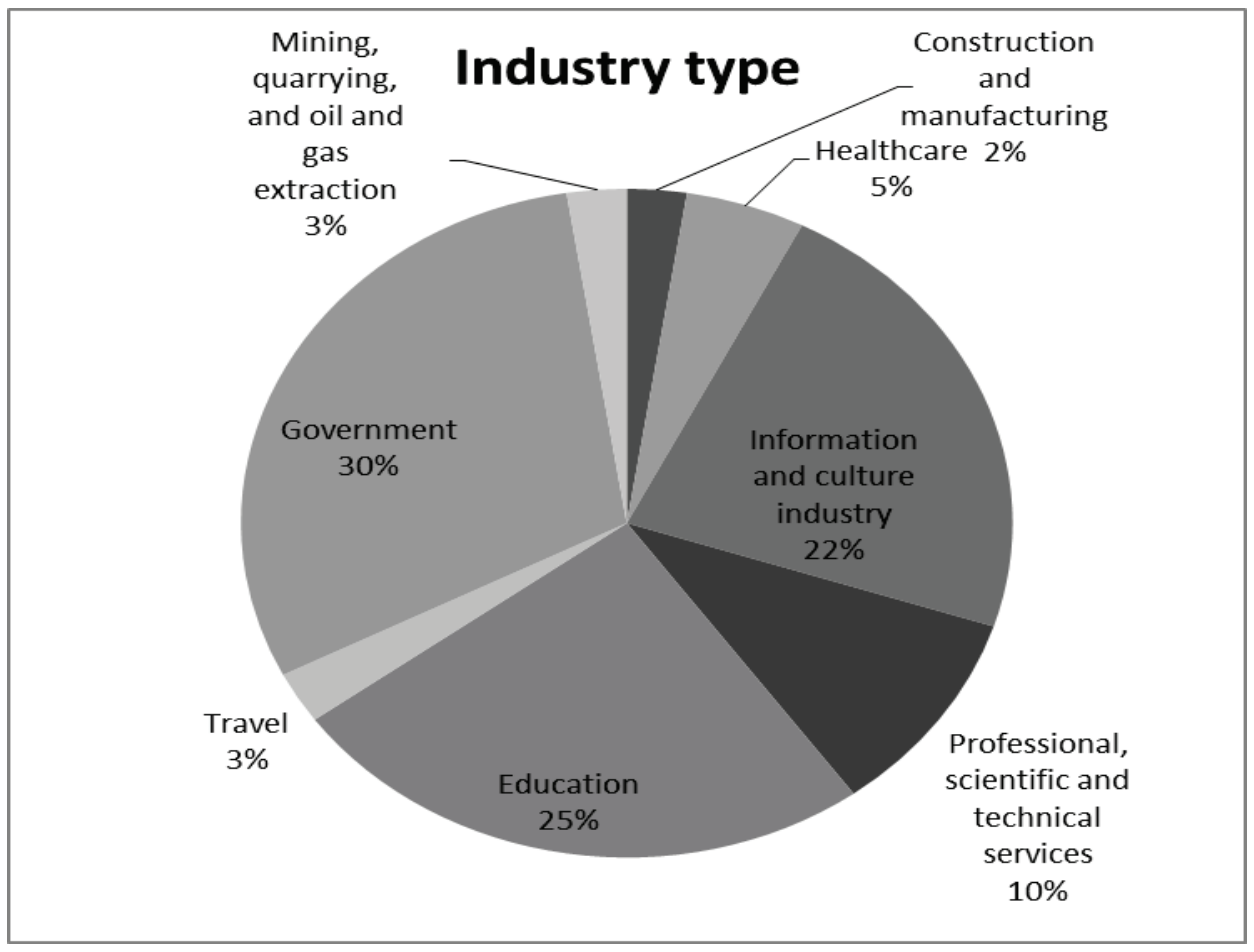

**Figure 2: Survey participant by industry type (n=40)**

The types of cloud deployment model reported by the participants are depicted in Figure 3.

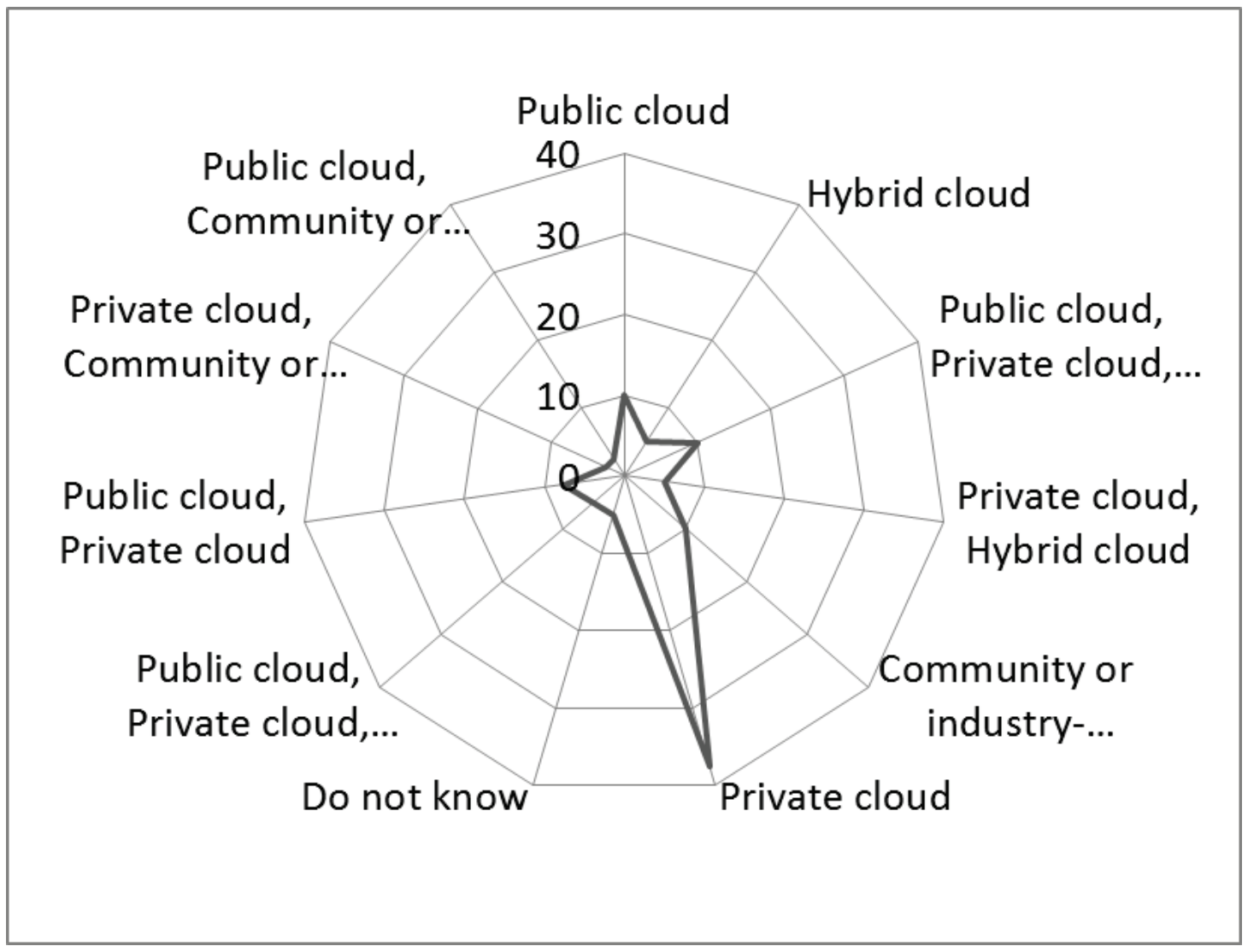

**Figure 3: Cloud deployment model reported by survey participants (n=40)**

Slightly more than a third of the participants (and they are generally from government agencies and educational institutions) reported that their organisation is deploying private cloud (37.5%). A small number of participants (7.5%) reported that their organisations used more than one cloud models (i.e. both private and public cloud services are used by different departments within the organisation). Figure 4 shows the breakdown of the cloud architecture in the participant's organisation.

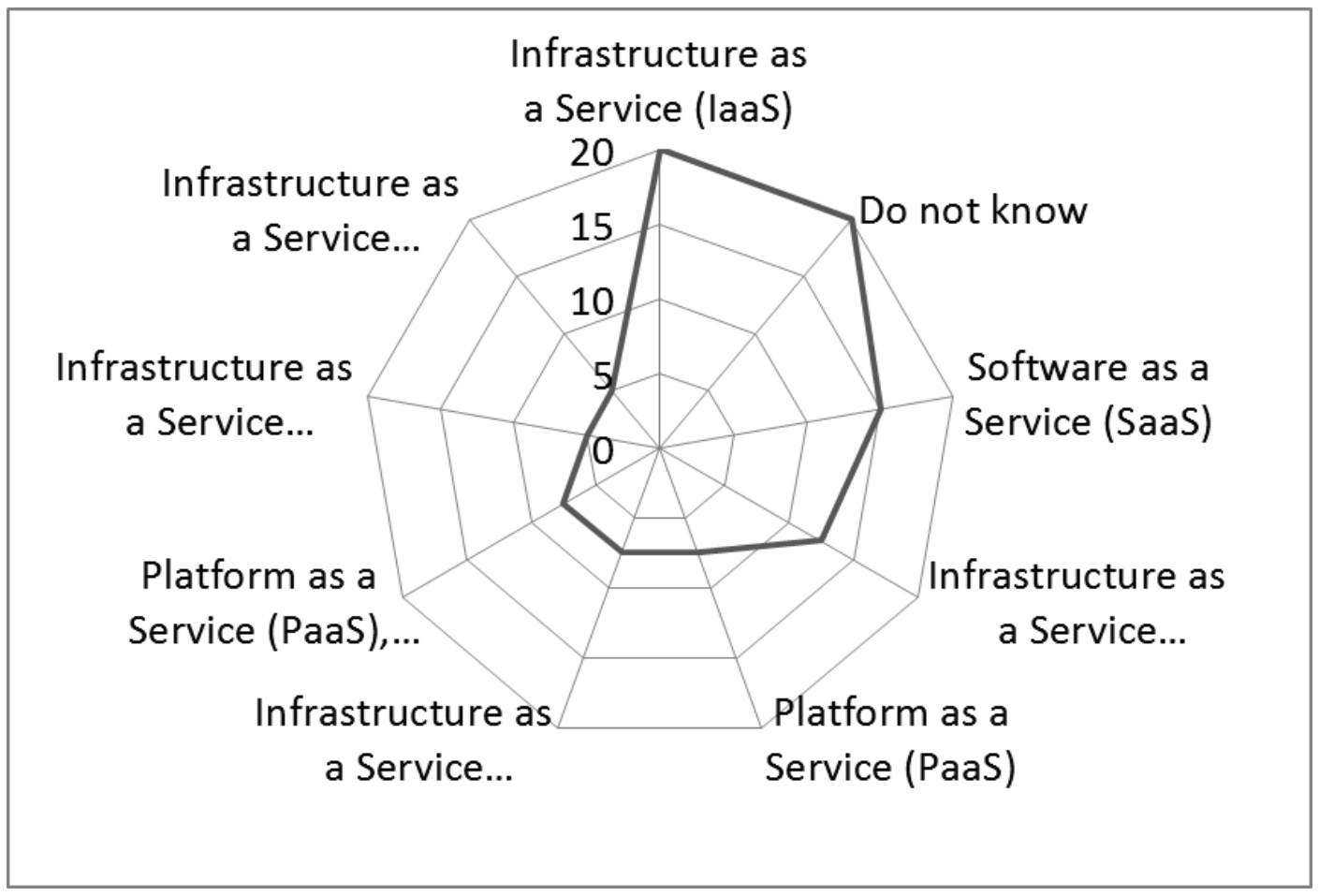

**Figure 4: Cloud architecture reported by survey participants (n=40)**

Infrastructure as a Service (IaaS) is reportedly the most deployed architecture (~20%), although this number may be slightly higher as one of five participants reported that they did not know which cloud architecture are being used in their organisation.

## *Construct Validity and Reliability Analysis*

Construct validity of the scale for measuring the independent variables was examined using Exploratory Factor Analysis (EFA). EFA groups variables into a smaller set to facilitate analysis and interpretation.

We used the Kaise-Mayer-Olkin (KMO) and Bartlett's test to analyse the survey responses. The KMO overall measure of sampling adequacy was 0.673, and the Bartlett's test of sphericity was significant ($p < .05$), indicating that the data is adequate and can be used for factor analysis. Using principle component analysis with Varimax rotation, the analysis generated a 5-factor solution, indicating approximately 80% of the total variance (eigenvalues greater than 1; see Table 4). A total of 16 items were loaded cleanly on the expected factors while the remaining items from Table 2 were omitted due to a low factor loading (<0.5).

| Factors | Item text | Factor loading | Eigenvalues | % of Variance |
|---|---|---|---|---|
| (1) Perceived vulnerability ($\alpha = 0.920$) | In your view, what is the likelihood of the following cloud specific threats faced by your organisation? | | 5.125 | 32.034 |
| | Cloud service abuse by outsider | 0.842 | | |
| | Denial of service | 0.813 | | |
| | Data loss by outsider | 0.807 | | |
| | Inadequate due diligence | 0.799 | | |
| | Account or service traffic hijacking by outsider | 0.785 | | |
| | Insecure software interfaces | 0.769 | | |
| | Data loss by insider | 0.732 | | |
| | Exploit people | 0.718 | | |
| (2) Response cost (α = 0.914) | What are the barriers to an effective information security incident handling strategy implementation at your organisation? | | | |
| | Lack of support from top management | 0.954 | 2.049 | 12.804 |
| | Lack of budget | 0.932 | | |
| (3) Perceived Severity (α = 0.880) | What are the consequences of an ineffective information security incident handling policy to your organisation (i.e. cloud service user)? | | 1.992 | 12.449 |
| | Financial loss | 0.918 | | |
| | Legal implications | 0.922 | | |
| (4) Self-efficacy (α = 0.810) | My organisation provides adequate budget to handle information security incident. | 0.900 | 1.915 | 11.968 |
| | My organisation provides adequate in-house staff to handle information security incident. | 0.761 | | |
| (5) Response efficacy (α = 0.773) | The policy is adequate to handle information security incident(s) on cloud environment efficiently. | 0.895 | 1.733 | 10.833 |
| | The workshop or seminar increases my awareness on information security and/or incident handling. | 0.813 | | |
| | **Cumulative % of Variance Values** | | | **80.087** |

**Table 4: Factor Loading based on Principle Component Analysis with Varimax (n=40)**

Cronbach's alpha test was carried out for reliability analysis of the items in the constructs. In interpreting the alpha (α) value, items that exhibit α >.70 indicate good reliability of the constructs (Nunally 1978). The data shows that the overall reliability of Cronbach's alpha reflects a high level of internal consistency (see Table 4). It can, therefore, be concluded that items used are valid and sufficiently reliable. We then proceed with the hypotheses testing using a Pearson Correlation Coefficient.

## *Hypotheses Testing*

When the finalised constructs indicate acceptable validity and reliability, it is necessary to measure the strength and direction between the incident handling strategy adoption and each factor. A Pearson correlation was conducted to determine the statistical association between incident handling strategy adoption (IHSA) and the five PMT factors used in our model (PVUL, SEF, REF, RCO, and PSEV). Table 5 presents the results of the bivariate correlation between the five factors and the test variable (IHSA).

As previously hypothesised, PVUL did significantly affect IHSA at the 0.01 level (r=.424; n=40; p<.001) with a relatively moderate positive correlation. Thus, H1 was supported. Consistent with our expectations, SEF and IHSA were positively significant at the 0.01 level (r=.456; n=40; p<.001). The data analysis also reported that REF had a highly significant effect on IHSA at the 0.01 level (r=.564; n=40; p<.0001). Accordingly, H2 and H3 were supported. Surprisingly, RCO had an insignificant effect on IHSA (r=-0.112;

n=40; p=ns). Hence, H4 was not supported. Lastly, PSEV affected IHSA at the 0.05 level (r=0.334; n=40; p<.05) and, therefore, H5 was supported.

| Factor | IHSA | PVUL | SEF | REF | RCO | PSEV |
|---|---|---|---|---|---|---|
| IHSA | — | | | | | |
| PVUL | .424** | — | | | | |
| SEF | .456** | .298 | — | | | |
| REF | .564** | .003 | .262 | — | | |
| RCO | -.112 | .211 | -.081 | -.123 | — | |
| PSEV | .343* | .286 | .000 | .148 | -.043 | — |
| Mean | 1.83 | 3.13 | 2.50 | 2.17 | 3.08 | 1.85 |
| SD | .712 | 1.265 | .987 | .844 | 1.118 | .834 |
| ** Correlation is significant at the 0.01 level (2-tailed)<br>* Correlation is significant at the 0.05 level (2-tailed) | | | | | | |

**Table 5: Correlation coefficients and descriptive statistic (n=40)**

We then carried out a linear multiple regression analysis to predict IHSA based on a combination of PMT factors as independent variables. Collinearity diagnostic results show that multicollinearity for this regression is not a major concern. Tolerance scores were all above 0.01 with the lowest score being 0.810, and VIF scores were all below 10 with a highest score of 1.319. A significant regression equation was found ($F (5, 34) = 9.245$, $p < .0001$). The overall model fit was $R^2 = .576$ (i.e. 57.6% of the variance in the global IHSA values). PVUL and REF were the only two variables in the model having a statistically significant effect on IHSA – see Table 6.

| Path | β | t-value | Result |
|---|---|---|---|
| PVUL | .323 | 2.518 | Supported |
| SEF | .230 | 1.857 | Not-supported |
| REF | .465 | 3.926 | Supported |
| RCO | -.096 | -.826 | Not-supported |
| PSEV | .178 | 1.488 | Not-supported |

**Table 6: Multiple Regression Model Predicting Incident Handling Strategy Adoption (n=40)**

## Discussion

Our study found that perceived vulnerability to have a significant effect on the adoption of a cloud incident handling strategy in the participant organisations. Our findings from the survey and the interviews echoed findings from previous studies, such as Ifinedo (2012) and Ng et al. (2009), which had found that perceived vulnerability influenced user behaviour. One of the interviewees (I1) also explained that his organisation opted for private cloud deployment due to concerns regarding the security and privacy of the organisational data and the uncertainty in the trustworthiness of the public CSP. For public CSUs who had to rely on their CSP's incident handling strategy, the organisations would need to conduct appropriate due diligence (e.g. thoroughly examine their organisation's and CSP's security requirements) before migrating to the cloud service. Another interviewee (I2) explained that:

> *I2: "Before we migrated to the X system, we had to complete all sorts of security checks as well as conducting other due diligence measures" [Translated by authors]*

The awareness at the Perception level forms the knowledge basis required to proceed to the next level.

**Proposition 1**: The more informed CSUs are regarding security and privacy risks associated with the use of the cloud, the greater their motivation to implement incident handling strategy within the organisation will be.

Our results also suggest that Self-Efficacy and Response Efficacy at the Comprehension level significantly influenced the adoption of incident handling strategy, which are consistent with findings from previously published empirical studies such as in Siponen et al. (2014) and Vance et al. (2012). The importance of self-efficacy reported by the survey participants is, perhaps, due to the high number of private CSUs in our study. Generally, a reputable CSP known to have a reputation for complying with international security and privacy governance protocols can increase the user's confidence and trust in the CSP (I2, I4). For instance, interview participant I4 reported that:

> I4: *"Due to the reputation of the CSP, we trust our data with them as we believe that they will ensure the security and privacy of our data." [Translated by authors]*

Interview participant I3 explains how their organisation can benefit by outsourcing their incident handling to a third-party vendor, such as a managed security service provider. Such an option may be more suited for SME-type organisation since it would not normally have the resources to adopt an effective incident handling strategy.

> I3: *"Having an incident handling strategy will be useful for Managed Security Services (MSS) providers, as clients don't need to know in detail the internal process of the vendor. This could result in significant savings, in terms of costs incident response time, and other resources." [Translated by authors]*

Our findings did not indicate that Response Cost had a significant influence on the adoption of incident handling strategy. Similar observation was also reported in previous research, such as those of Ifinedo (2012) and Meso et al. (2013). Interview participant I2, when asked about barriers encountered during the implementation of incident handling strategy in his organisation, explained that:

> I2: *"No, I don't think we have encountered any barriers. We are fine as we haven't had any security incidents… So I think it should be quite straight forward to report an incident if we have one. " [Translated by authors]*

More than half of the survey participants (60%) also reported that they were not aware of any cloud-related incident. This could be due to the fact that incidents were undetected or unreported.

At the Comprehension level, our findings indicated that the participants understand the benefits of implementing an incident handling strategy to minimise security risks, and are able to assess their capability in undertaking the strategy.

It also appeared that perceived severity influenced the adoption of incident handling strategy within the organisation, and similar findings are reported by Siponen et al. (2014) and Ifinedo (2012). As an example of an ineffective strategy, weak enforcement of authentication policy would likely increase the potential of a data breach incident and, consequently, affect the confidentiality and integrity of the organisational data, as well as resulting in significant reputational and financial damages. In addition, the cost involved in post-incident damage control should be taken into consideration by the organisation to ensure that there is an effective strategy, as described by one interview participant.

> I4: *"If we do not have an incident handling strategy, a major security breach would require a significant effort and cost in the damage control exercise, and in our Internet-connected organisation, any security breaches resulting in compromise of company data will have a significant*

> *long-term impact for us. For example, the high profile iCloud security breach taught us a lesson. Even if we managed to tighten our security system after the incident has been detected, there is absolutely nothing we can do to recover the stolen data or prevent the dissemination of the stolen data." [Translated by authors]*

This study, therefore, proposes the following propositions,

**Proposition 2 (a):** CSUs are more likely to adopt incident handling strategy when they are confident in their capability to implement the strategy.

**Proposition 2 (b):** CSUs are more likely to adopt incident handling strategy when they understand the benefits of undertaking the strategy.

**Proposition 2 (c)**: CSUs are more likely to adopt incident handling strategy when they can understand the severity of a situation due to an inefficient or ineffective strategy.

The regression analysis also determined that Perceived Vulnerability and Response Efficacy were significant factors in influencing the adoption of an incident handling strategy by the participants. We, therefore, suggest that a successful adoption of an incident handling strategy by an organisational CSU involves the nexus between Perceived Vulnerability, Self-Efficacy, Response Efficacy, and Perceived Severity.

**Proposition 3**: Overall, CSU's adoption of incident handling strategy is influenced by their perceived vulnerability of cloud security, perceived severity of cloud incidents, and the self-efficacy and benefits of the adopting and implementing the strategy.

## Conclusion and Future Work

This paper highlighted the importance of having an incident handling strategy for organisational cloud service users, particularly to ensure the security and privacy of their organisational data. We proposed a conceptual model (an integration of the Situation Awareness model and Protection Motivation Theory) to guide us in identifying influencing factors in the adoption of an incident handling strategy by organisational cloud service users. Based on our survey and interviews of participants from 40 different organisations in Malaysia, we determined that Perceived Vulnerability, Self-Efficacy, Response Efficacy, and Perceived Severity are factors that have a significant influence on the adoption of an incident handling strategy by organisation cloud service users.

However, the responses were small in number, by comparison to the number of organisational cloud service users in Malaysia; therefore, the findings may not be generalisable. Future work would include conducting the study on a larger scale within the country, which would provide a statistically sound national data. Future study would also include the refinement of the measurement items (e.g. considering cloud deployment and service models as the moderator or control variables ) and the conceptual model (e.g. by integrating constructs and variables from other theoretical models) so that the model is more robust.

## Acknowledgements

The first author is currently a PhD student at the University of South Australia, supported by Ministry of Education, Malaysia (MOE) and University of Tun Hussein Onn Malaysia (UTHM). The views and opinions expressed in this article are those of the authors alone and not the organisations with whom the authors are or have been associated and supported. The authors would also like to thank the anonymous reviewers for providing constructive and generous feedback. Despite their invaluable assistance, any errors remaining in this paper are solely attributed to the authors.